\documentclass{PoS}
\def\ktresummation{$k_t$-resummation}
\def\ee{\end{equation}}
\def\be{\begin{equation}}
\def\eea{\end{eqnarray}}
\def\bea{\begin{eqnarray}}
\def\pp{$pp$}
\def\pbarp{${\bar p}p$}

\def\sigtot{\sigma_{tot}}

\def\x{cross-section}

\def\vecb{{\vec b}}
\def\vecq{{\vec q}}

\def\nbar{{\bar n}}

\title{Soft gluon resummation in the infrared region and the Froissart
bound}

\ShortTitle{Soft gluon resummation in the infrared region and the Froissart
bound}

\author{\speaker{Giulia Pancheri} \thanks{Visitor at CTP, MIT, Cambridge, MA, USA}\\
INFN Frascati National Laboratories, Via E. Fermi 40, I00044 Frascati, Italy\\
E-mail: \email{pancheri@lnf.infn.it}}
\author{Agnes Grau\\
Departamento de Fisica Teorica y del Cosmos, Universidad de Granada, Spain\\
      E-mail: \email{igrau@ugr.es}}
      \author{Rohini M. Godbole\\
      Centre for High Energy Physics, Indian Institute of Science, Bangalore, 560012, India \\
      Email: \email{rohini@cts.iisc.ernet.in}}
\author{Yogendra N. Srivastava\\
INFN and University of Perugia, Perugia I06123, Italy      \\
E-mail: \email{yogendra.srivastava@pg.infn.it}}

\abstract{We describe the taming effect induced by soft gluon $k_t$-resummation
on the rapid rise of  QCD mini-jet contributions to the total cross-sections.
We use an eikonal model in which the rise of the total cross-section is due 
to mini-jet contribution and perform our calculation with current Parton 
Density Functions (PDFs). The impact parameter distribution we use is obtained 
as the Fourier transform of the resummed $k_t$-distribution of soft gluons 
emitted from the initial state during the collision. The emission, which is 
energy dependent, destroys the initial collinearity of partons. In this model, 
the strong power-like rise due to the increasing number of low-x gluon 
collisions is tamed by the acollinearity induced by soft gluon kt-resummation 
down to zero gluon momenta. It   explicitly  links a singular soft gluon 
coupling in the infrared region to the behaviour dictated by the Froissart 
bound for the total cross-section.  The model describes well both proton and 
photon processes at the current accelerator energies and gives predictions for
the TeV range.  For photons the model predictions are also in agreement with 
the fits by Block and Halzen, obtained using Finite Energy Sum Rules (FESR) at 
low energy and an asymptotic behaviour consistent with the Froissart bound.}

\FullConference{XVIII International Workshop on Deep-Inelastic Scattering and Related Subjects, DIS 2010\\
		April 19-23, 2010\\
		Firenze, Italy}

\begin{document}

\section{Our model}
Our  model for the total cross-section \cite{our2005} starts with the simplified eikonal formulation
\begin{equation}
\sigma_{total}\approx 2\int d^2{\bf b} [1-e^{-{\bar n} (b,s)/2}]
\end{equation}
where the real part of the scattering amplitude has been approximated to zero and  
 \begin{equation}
  {\bar n}  (b,s)={\bar n} _{soft}(b,s)+{\bar n}_ {QCD}(b,s)=A_{soft}(b,s)\sigma_{soft}(s)+A_{BN}(b,s) \sigma_{\rm jet} (s,p_{tmin})
  \end{equation}
The mini-jet \x , $\sigma_{\rm jet} (s,p_{tmin})$, 
is calculated using parton-parton cross-sections and  DGLAP evoluted PDFs, 
integrating over all parton final state momenta such that 
$p_t\ge p_{tmin}$. $A_{BN}$ is the impact parameter distribution for the 
mini-jet processes which we describe in the next section and which leads to a 
cut-off  at large $b$-values with a behaviour stronger than an exponential 
\cite{ourFroissart}. The soft part of the cross-section, which dominates up to 
$\sqrt{s}\sim  10\  GeV$ is parametrized as described in \cite{our2005}.

We shall refer to this model as  Bloch-Nordsieck (BN) model 
 because of the indispensible role played in our model by the 
resummation of soft massless quanta in gauge theories.
We were inspired to build our model by the classic work of  Bloch and Nordsieck \cite{Bloch:1937pw} in electrodynamics, where they first  pointed  out that 
only the emission of an infinite number of soft photons can lead to a finite 
result.

\begin{figure}[h]
\centering
\vspace{-10mm}
\begin{minipage}{.5\linewidth}
\includegraphics[width=1.\textwidth]{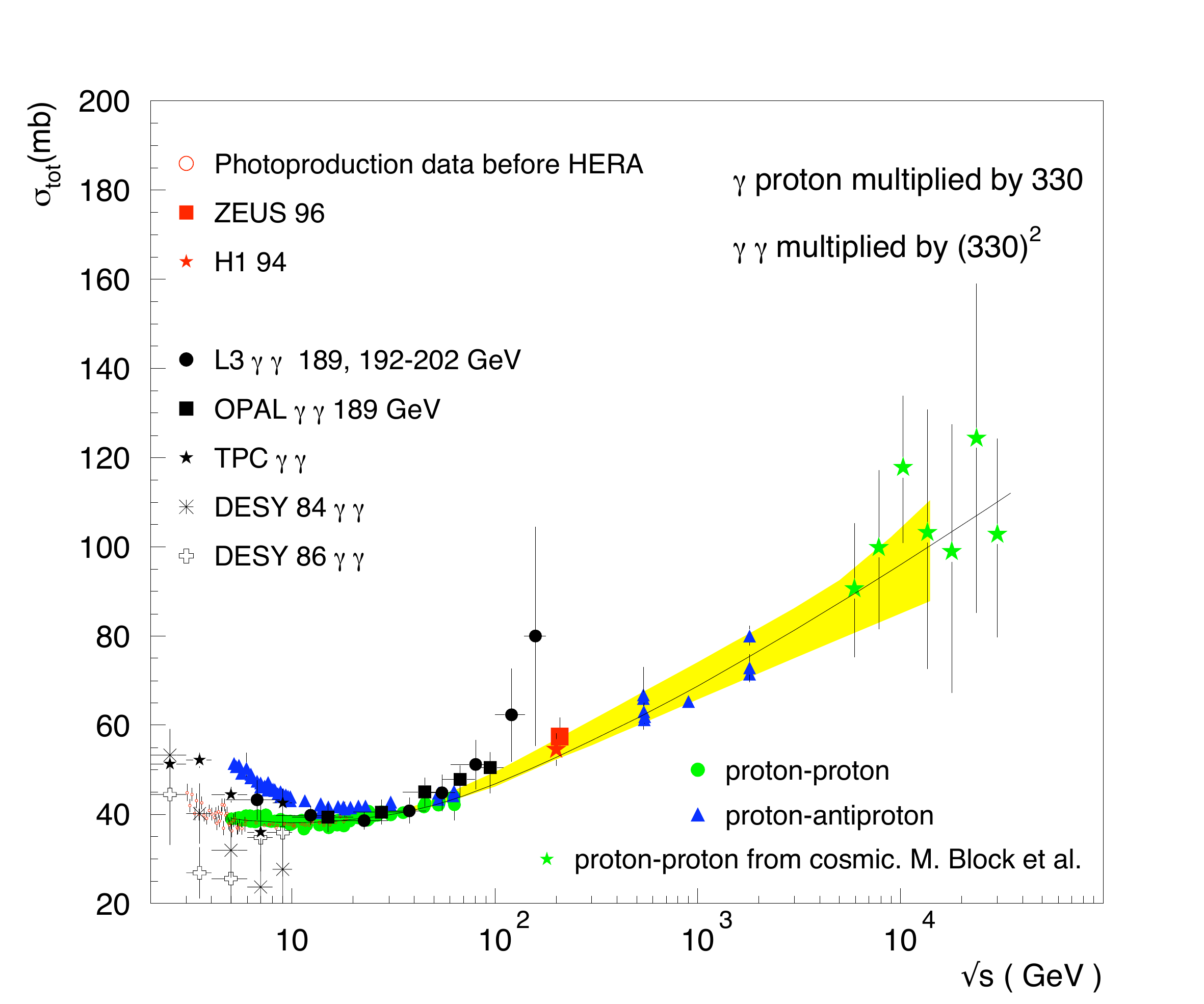}
\label{fig:bande}
\end{minipage}%
\begin{minipage}{.5\linewidth}
\includegraphics[width=1.\textwidth]{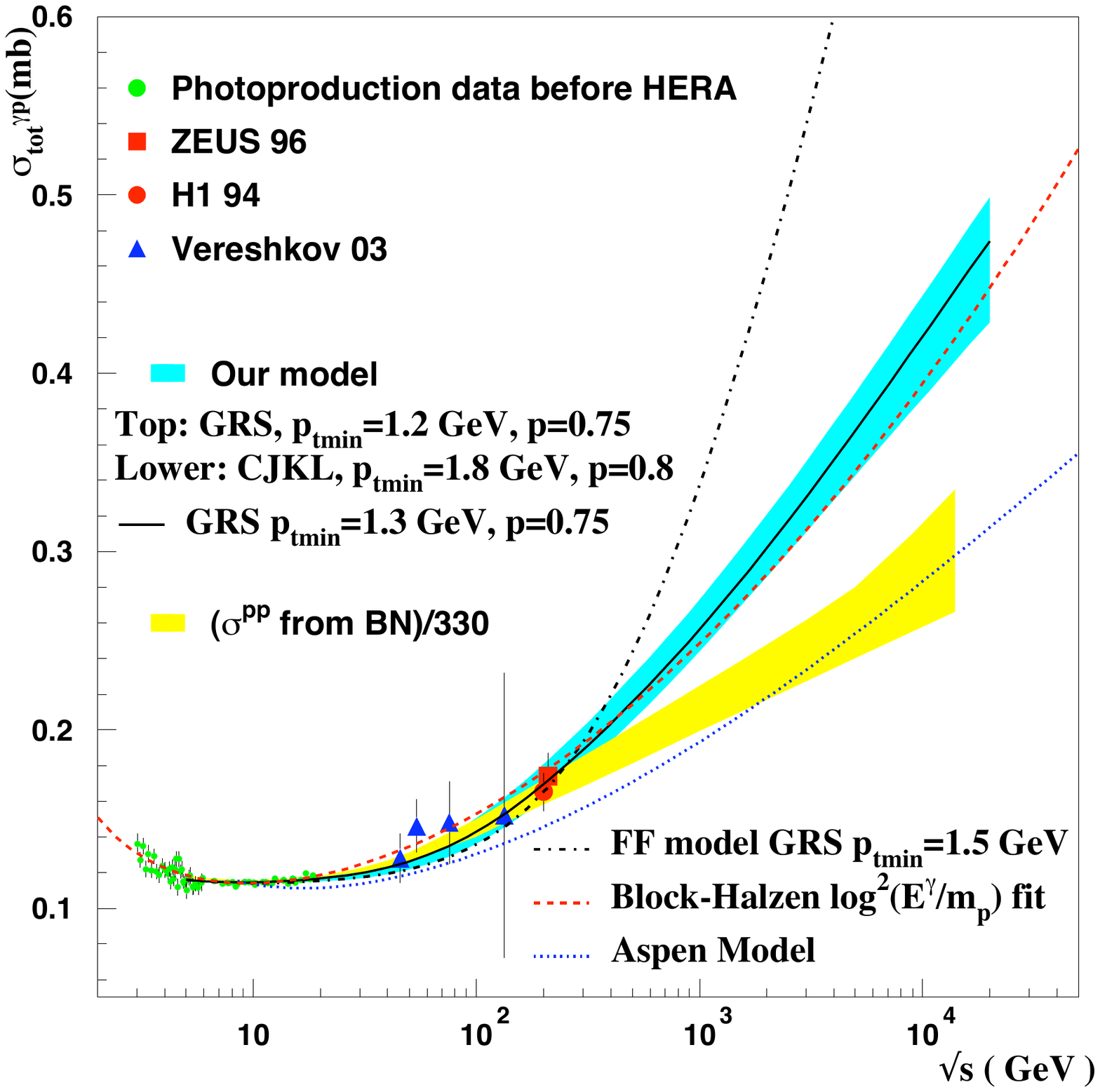}
\label{fig:gamp}
\end{minipage}
\vspace{-25mm}
\caption{From \protect\cite{ourEPJC}: at left is a compilation of data for 
$pp$, $p {\bar p}$, $\gamma p$ and $\gamma \gamma$  with the yellow band and 
central black line showing our predictions. Photon data have been multiplied 
by an {\it ad hoc factor} proportional to $\alpha_{QED}^{-1}$. At right is
the  $\gamma p$ total cross-section, with  the yellow band obtained by
multiplying  the \pp\ band of the left panel by a numerical factor, the blue 
band is our model for $\gamma p$ and the Block-Halzen line is from \cite{BH}.}
\end{figure}

In our model, such a soft gluon \ktresummation\ and its implementation down 
into the gluon infrared momentum region, constitute the mechanisms through 
which the fast initial rise of all the total cross-sections is transformed into
a smooth logarithmic rise, consistent with the Froissart bound.
We have obtained results  for proton and photon total cross-sections as shown in Fig.~\ref{fig:bande} from \cite{ourEPJC} . 
In Fig.~\ref{fig:bande}, left panel,  we show a compilation of proton and 
photon total cross-section data, where photon data have been multiplied by an 
{\it ad hoc } factor proportional to $\alpha_{QED}^{-1}$. Our model description
for \pp  \ and \pbarp \ is indicated by the band and by the full line. 

When dealing with photons, the eikonal model we have used needs to be modified 
so as to take into account the transition of the photon into a hadron  when the
photon interacts with matter. Indicating the probability for this transition as
$P_{had}$, we follow the model proposed originally in \cite{fletcherhalzen} 
and write
\begin{equation}
\sigma^{\gamma p}_{total}= 2P_{had} \int d^2{\bf b} [1-e^{-{\bar n}^{\gamma p} (b,s)/2}]
\end{equation}
with the average number of collisions to be calculated using again minijets and
the known PDFs for the photon. With  $P_{had}=1/240$ (through  Vector Meson 
Dominance arguments), for $\gamma p$ we find   \cite{ourEPJC} that the  \x\ 
remains proportional to the proton \x s up to present accelerator energies 
(yellow band in right panel of Fig.~\ref{fig:bande}), but it becomes quite a 
bit larger at higher energies. On the other hand, our results agree with the 
results of Block and Halzen  \cite{BH}, who combine a maximally allowed 
$\ln^2{s}$ behaviour at high energy with a low energy fit from  the FESR.

\section{The impact parameter distribution in the BN model}
The impact parameter distribution $A(b)$ has a simple semi-classical 
interpretation as the  convolution of form factors of the colliding 
particles \cite{DurandPi}. However, in the energy region where mini-jets start 
playing a role, the form factor expression is inadequate. Our proposal is that 
the impact parameter distribution $A_{BN}(b,s)$ in parton-parton scattering be 
derived from initial state soft gluon \ktresummation \ \cite{DDT,PP}. It 
is given by the normalized Fourier transform of the $k_t$-resummed expression 
for soft gluon initial state emission in parton-parton collision.  Because the 
total \x \ is dominated by large impact parameter values, we have revisited 
this distribution, extending the integration  over single soft gluon momenta 
into  the infrared region, as  discussed in  \cite{Corsetti} and described at 
length for protons  \cite{our2005}  and recently for  photons \cite{ourEPJC}.

 The expression we use 
 is
\be
 A_{BN} (b,s)
 =\frac {
  e^{-h(b,q_{max})}
  }
{
\int d^2{\vecb}\   e^
{
-h(b,q_{max})
}
}=
\frac
{
e^
{
-\frac{16}{3\pi}\int_0^{qmax}
 \frac{dk_t}{k_t}
  \alpha_{eff}(k_t)\ln 
(\frac{2q_{max}}{
k_t})
[1-J_0(b,k_t)]}
}
{\int d^2{\vecb}\   e^{-h(b,q_{max})}}
\ee
where the function  $h(b,q_{max}) $
is obtained from soft gluon resummation techniques
 and has a logarithmic  energy dependence through the scale $q_{max}$, which is proportional to $p_{tmin}$.  To obtain the taming effect which will change the fast rise of mini-jets in a logarithmic behaviour {\it a' la Froissart}, it is necessary to extend  resummation down to single soft gluon momenta below $\Lambda_{QCD}$. To do this, a model for the infrared gluon distribution has to be introduced, as explained in \cite{ourFroissart}. With an effective coupling 
$\alpha_{eff}(k_t)\sim k_t^{-2p}$ for the single soft gluon transverse momentum  in the region $0 \le k_t \le \Lambda_{QCD}$
and $1/2<p<1$, we have shown in \cite{ourFroissart} how this expression for $A_{BN} (b,s)$ introduces a strong cut-off in $b$-space and changes the violent rise of mini-jets into a smooth behavior in the total \x\ . Namely, we found that $\sigtot \sim (\ln s)^{1/p}$, a behaviour consistent with the limitations imposed by the Froissart bound.   

For the low energy region, where minijets do not yet play a role, 
we use the parametrization   proposed in \cite{our2005} for $A_{soft}(b,s)$, which was also inspired by the soft gluon resummation model.
\begin{figure}[p]
\centering
\begin{minipage}{.45\linewidth}
\includegraphics[width=.85\textwidth]{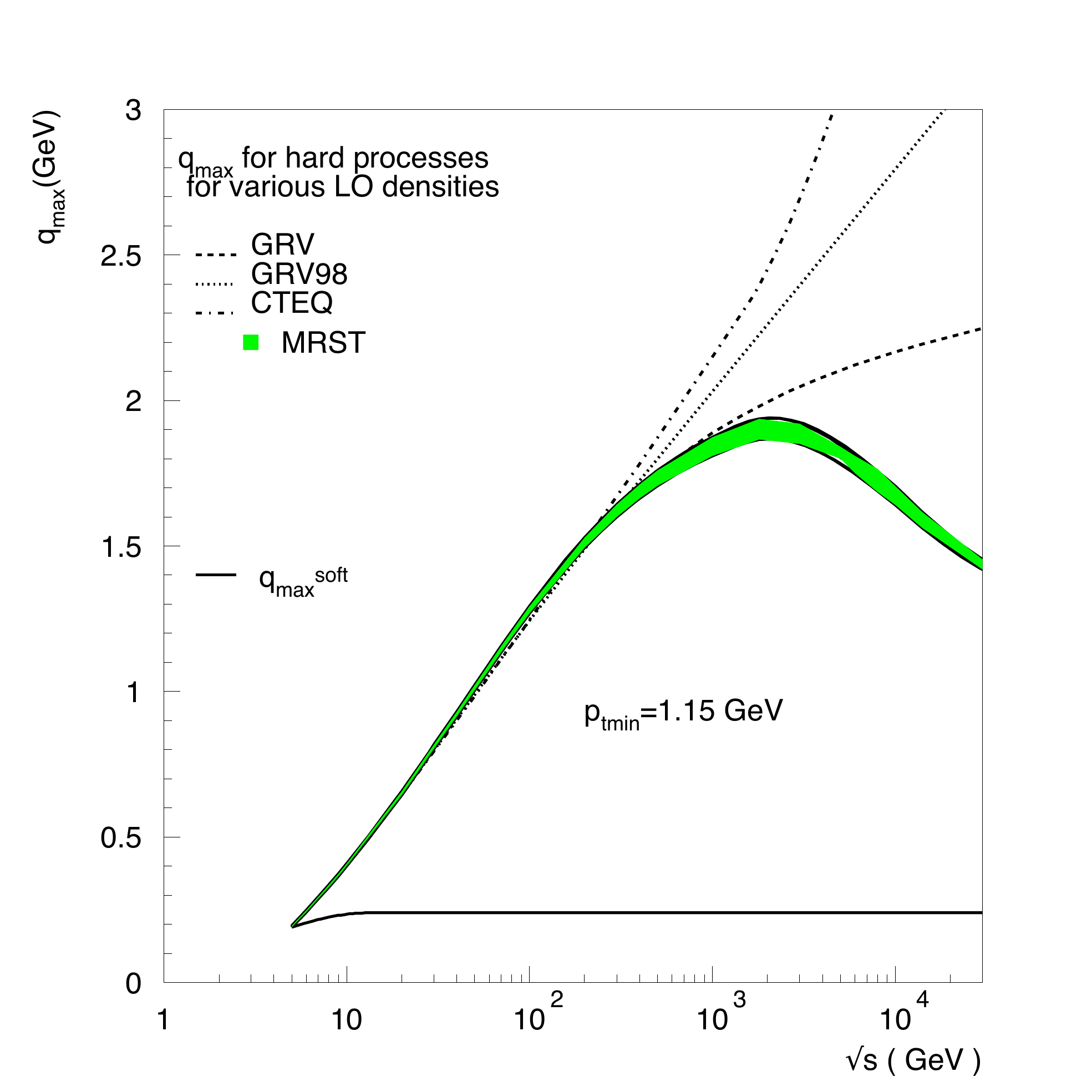}
\caption{The maximum single soft gluon transverse momentum \pp\  \ or \pbarp  \ scattering for different PDFs.}
\label{fig:qmax_115_band}
\end{minipage}
\hspace{1cm}
\begin{minipage}{.45\linewidth}
\includegraphics[width=.85\textwidth]{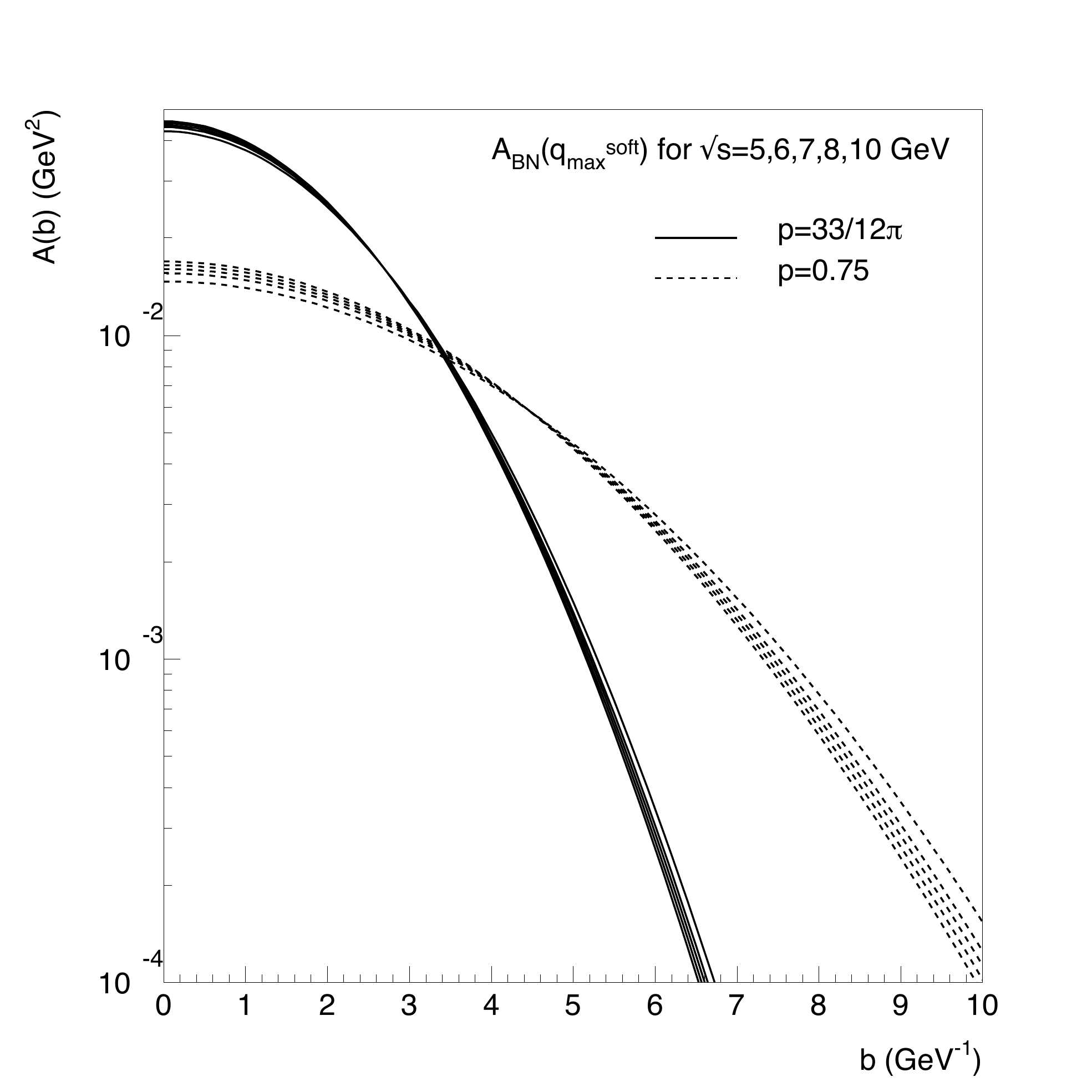}
\caption{The model impact parameter distribution  at various  low-energy values for two values of the parameter  $p$.
 }
\label{fig:abnsoft}
\end{minipage}
\end{figure}

\begin{figure}[h]
\centering
\begin{minipage}{.45\linewidth}
\includegraphics[width=.85\textwidth]{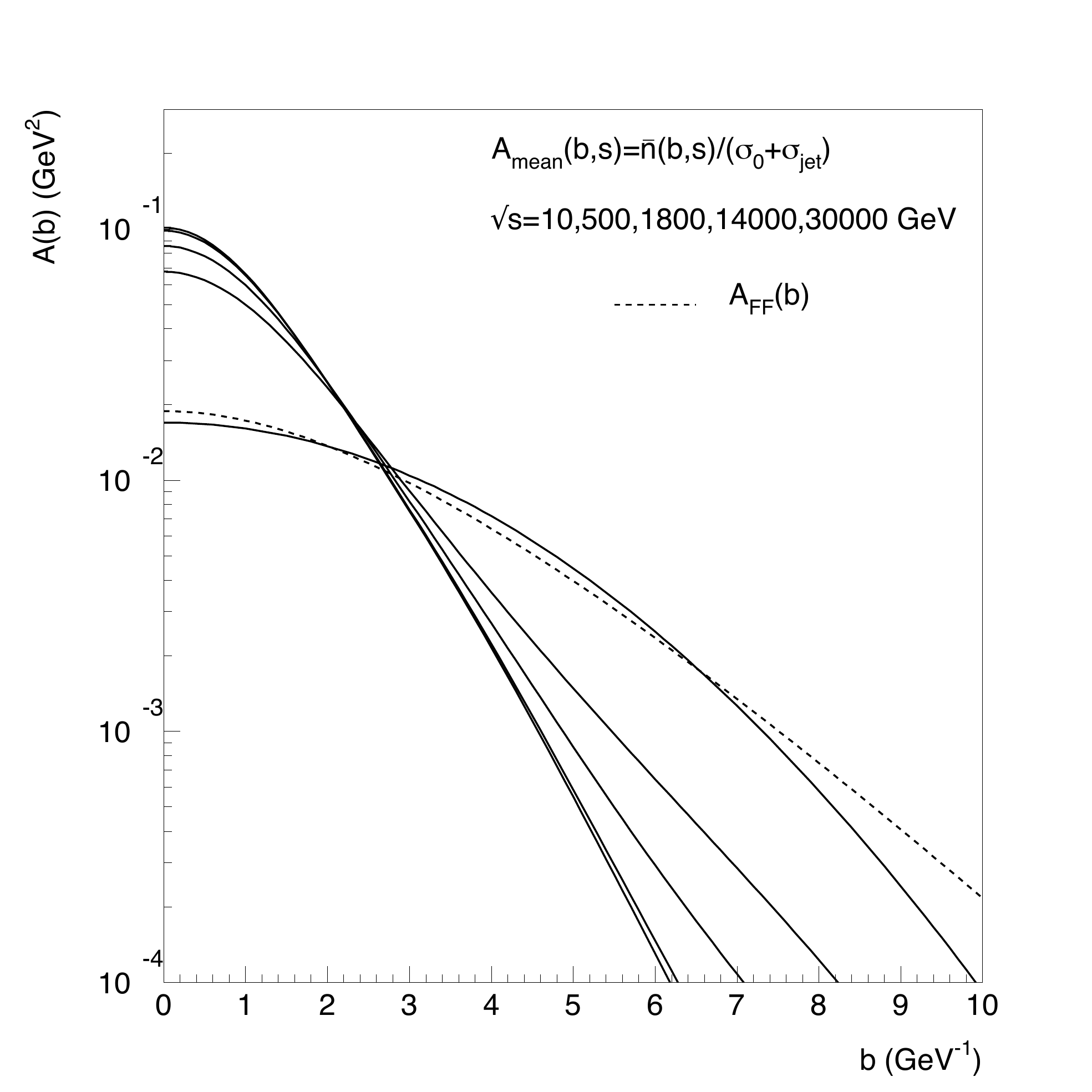}
\caption{The model impact parameter distribution  for protons 
 at various energies.}
\label{fig:abn_mean_p075}
\end{minipage}%
\hspace{1cm}%
\begin{minipage}{.45\linewidth}
\includegraphics[width=.85\textwidth]{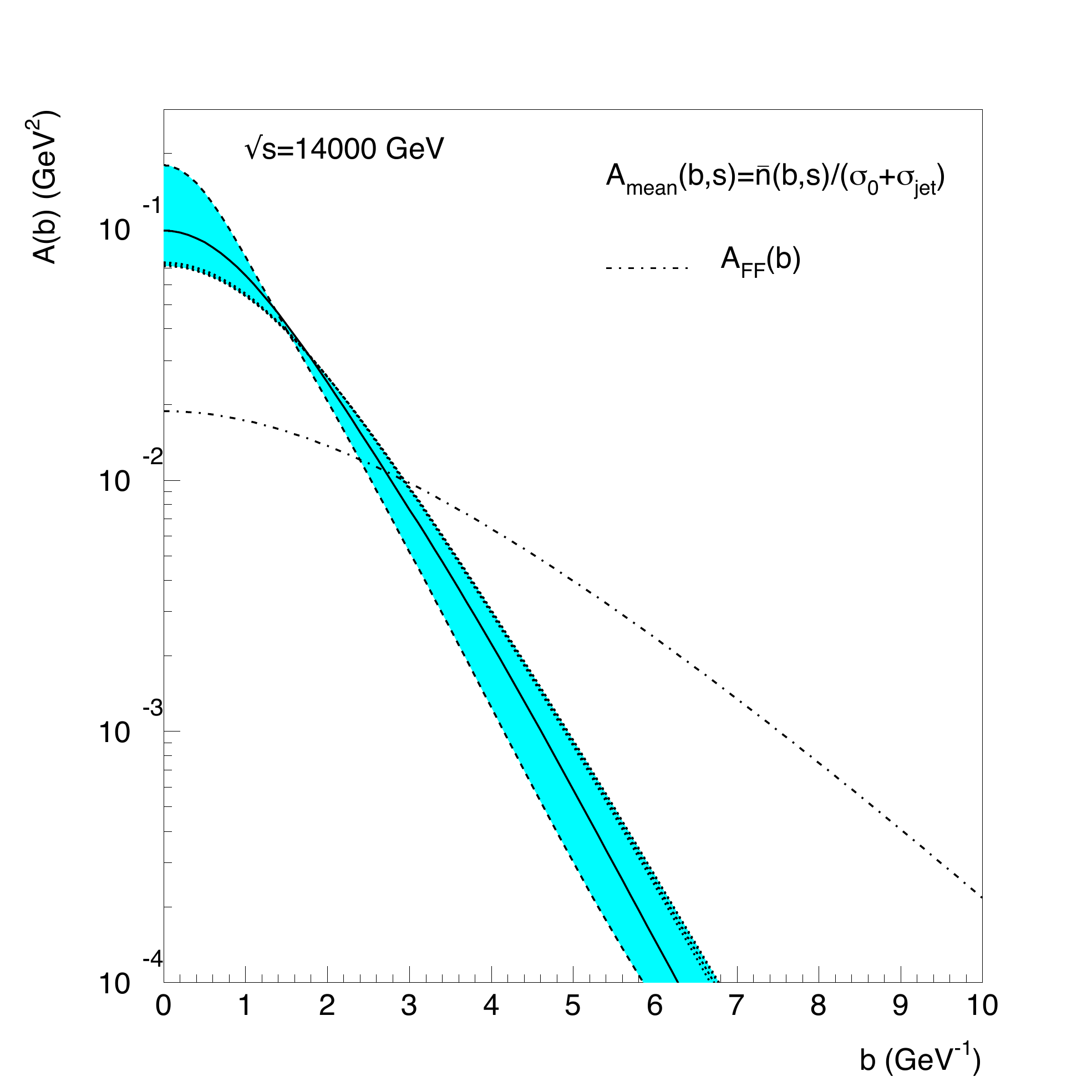}
\caption{The model
impact parameter distribution at LHC  for different PDFs.}
\label{fig:abn_mean_area_lhc}
\end{minipage}
\end{figure}

\begin{figure}[h]
\centering
\begin{minipage}{.5\linewidth}
\includegraphics[width=.8\textwidth]{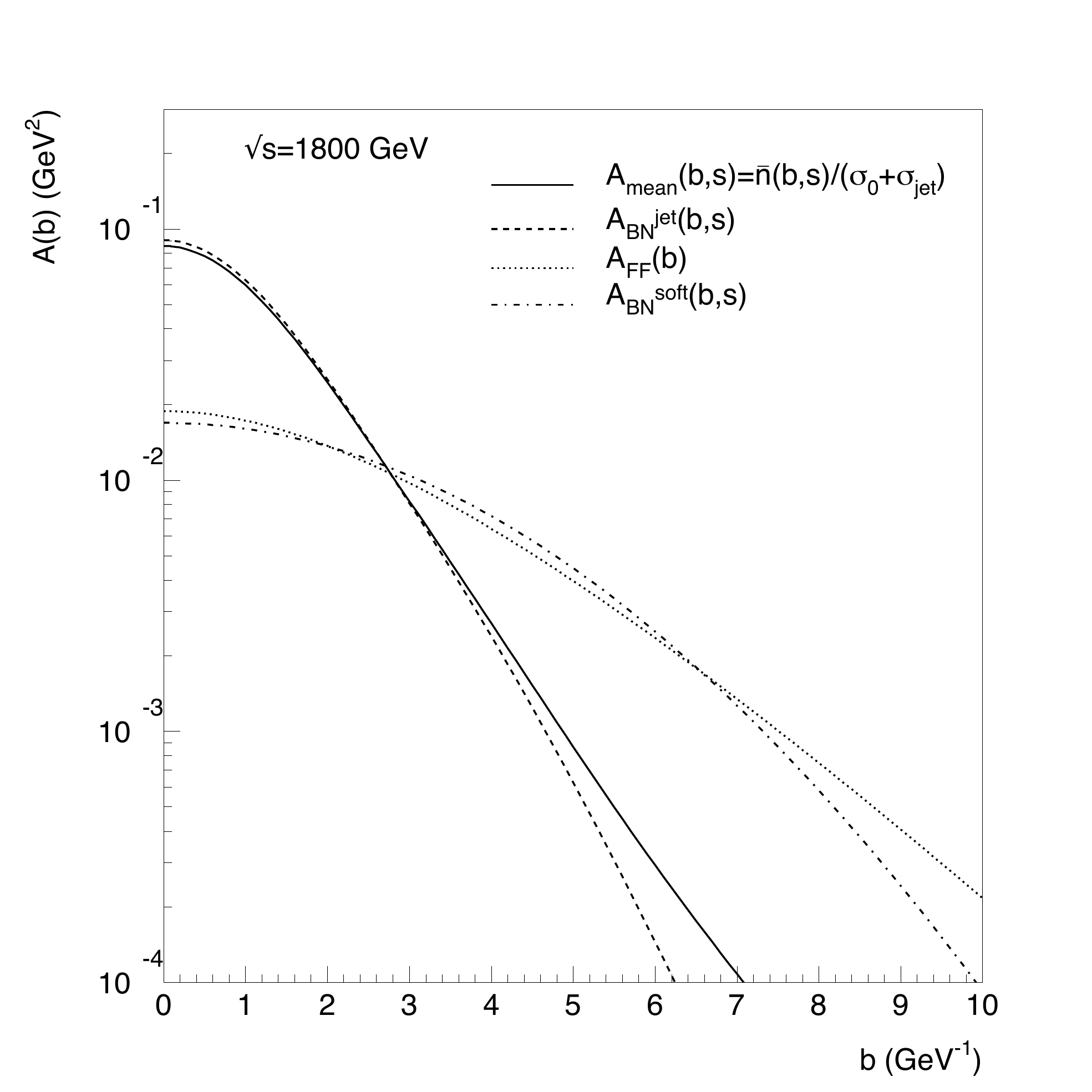}
\end{minipage}%
\begin{minipage}{.5\linewidth}
\includegraphics[width=.8\textwidth]{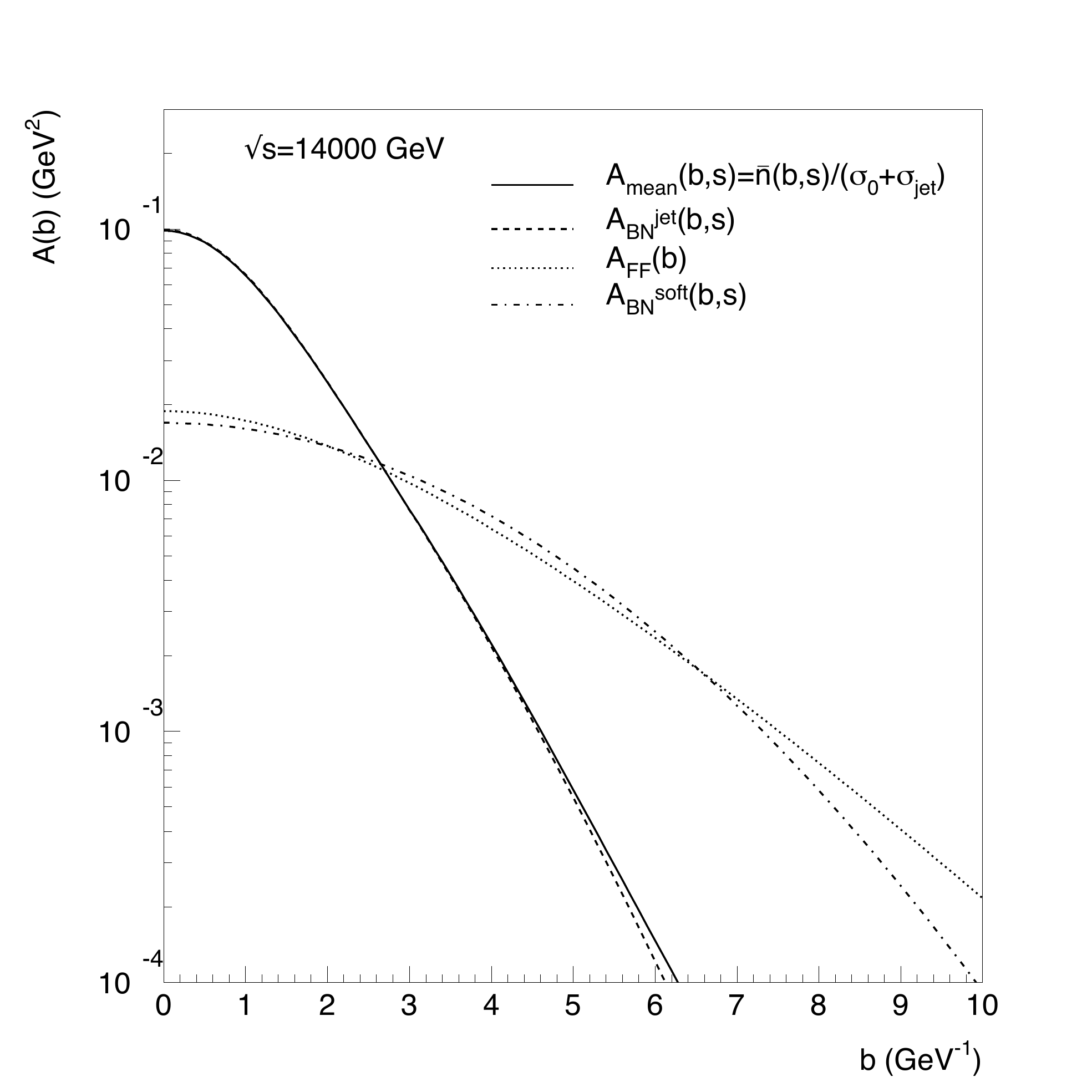}
\end{minipage}
\caption{Different  impact parameter distributions for the Tevatron and the LHC
energies.}
\label{fig:abn_mean_1800_14000}
\end{figure}
To compare our result for $\nbar ^{pp} (b,s)$ at LHC with other models, and in particular with models which use the  form factors both at low and at high energy,  we can define an average $b$-distribution such that 
\be
A_{mean}^{pp}(b,s)=\frac{\nbar (b,s)}{\sigma_{soft}^{pp}(s)+\sigma^{pp}_{\rm jet} (s,p_{tmin})}
\ee
We show the energy dependence of $q_{max}$ in Fig.~\ref{fig:qmax_115_band} and $A(b)$ from our \ktresummation \ model in  Figs.~\ref{fig:abnsoft},\ref{fig:abn_mean_p075},\ref{fig:abn_mean_area_lhc},\ref{fig:abn_mean_1800_14000}. In all the figures, we have chosen  $p_{tmin}=1.15\ GeV$, used the  GRV densities and $p=0.75$, unless indicated.  In these figures we  compare our model for $A(b,s)$ with the Form Factor model expression
\be
A_{FF}(b)=\frac{1}{(2\pi)^2}
\int d^2\vecq e^{i\vecq \cdot \vecb}[{\cal F}_p(q)]^2=\frac{1}{(2\pi)^2}
\int d^2\vecq e^{i\vecq \cdot \vecb}  \frac{1}{[1 + (q/\nu)^2]^4}
\ee
where ${\cal F}_p$ is the proton form factor 
with $\nu^2=0.71\ GeV^2$.

\acknowledgments
R.G.  acknowledges support from the Department of Science and
Technology, India under Grant No. SR/S2/JCB-64/2007, under the J.C.
Bose Fellowship scheme. Work partially supported by the Spanish MEC 
(FPA2006-05294 and FPA2008-04158-E/INFN) and by Junta de Andaluc\'\i a 
(FQM 101). G.P gratefully acknowledges the hospitality of the MIT Center 
for Theoretical Physics. YS is also an Emeritus Professor at Northeastern 
University Boston MASS and thanks the physics department for its hospitality.

\end{document}